\documentclass[epjST]{svjour}
\usepackage{graphics}
\usepackage[english]{babel}
\usepackage{mdframed,xcolor}
\usepackage{hyperref}
\hypersetup{
colorlinks=true,
citecolor=blue,
linkcolor=blue,
filecolor=blue,
urlcolor=blue}
\begin{document}
\title{The Emerging Energy Web}
\subtitle{}
\author{Marco Ajmone-Marsan\inst{1,2}\fnmsep\thanks{\email{marco.ajmone@polito.it}} \and David Arrowsmith  \inst{3}\fnmsep\thanks{\email{d.k.arrowsmith@qmul.ac.uk}}\and
Wolfgang Breymann \inst{4}\fnmsep\thanks{\email{bwlf@zhaw.ch}}\and
Oliver Fritz \inst{5}\fnmsep\thanks{\email{oliver.fritz@ch.abb.com}} \and Marcelo Masera \inst{6}\fnmsep\thanks{\email{marcelo.masera@ec.europa.eu}}\and Anna Mengolini \inst{6}\fnmsep\thanks{\email{anna.mengolini@ec.europa.eu}}\and Anna Carbone\inst{7,8}\fnmsep\thanks{\email{anna.carbone@polito.it}}}
\institute{Department of Electronic Network and Telecommunication, Politecnico di Torino, Italy \and IMDEA Network Institute, Madrid, Spain  \and Queen Mary University of London, UK \and Zurich University of Applied Sciences, Switzerland   \and ABB Switzerland Ltd, Corporate Research, Baden-D\"attwil, Switzerland \and JRC Petten, The Netherlands \and Department of Applied Science and Technology, Politecnico di Torino, Italy \and ETH Swiss Federal Institute of Technology, Zurich, Switzerland}
\abstract{There is a general need of elaborating energy-effective solutions for managing our increasingly dense interconnected world. The problem should be tackled in multiple dimensions -technology, society,  economics, law, regulations, and politics- at different temporal and spatial scales. Holistic approaches will enable technological solutions to be supported by socio-economic motivations, adequate incentive regulation to foster investment in green infrastructures coherently integrated with adequate energy provisioning schemes.
In this article,  an attempt is made to describe such multidisciplinary challenges with a coherent set of solutions to be identified to significantly impact the way our interconnected energy world is designed and operated.}
%end of abstract
%%
%
\maketitle
\clearpage
\tableofcontents

%\sloppy
\section{Introduction}
\label{intro}
Our society is urged by significant transformations of the energy system determined  by several factors.
The world energy demand is projected to increase by 50\% from 2005 to 2030, mainly due to the rapidly developing economies of the highly populated non-OECD  countries.
The growth of energy demand and carbon dioxide  emissions  together with the threat of climate change are key concerns for energy policy makers and society.
 \par
 Reductions of at least 50\% in global  carbon dioxide emission compared to 2000 levels  need to be achieved by 2050 to limit  the long-term global average temperature rise  between 2.0 and 3.0 degrees.  Another aspect of the same problem is  the continuous increase of the  energy cost obtained from fossil sources  due to the growing demand.  Fossil fuels (oil, gas, and coal) account for about 80\% of the global primary energy demand and  are supplied by relatively few producer countries, which have limited reservoirs.   Efforts to forge a long-term policy framework for tackling climate change and rising cost of fossil fuels are continuing (see Box 1 for a few examples).  The difficulty of reaching an agreement just based  on  top-down solutions  has been clearly demonstrated.  It was pointed out, on many occasions, that a concrete and overall transformation of our energy system can only happen in a concerted way with a bottom-up transformation at the elementary levels of the energy infrastructures with a reinforcing bidirectional feedback, where directives outlined at the top level, are taken up and reinforce initiatives already emerging at the bottom level \cite{MacKay,Chu,Tollefson,Peters,Wolsink,Jakob,Carvalho12}.
\par
As a result society at large, business and social actors defining the global energy landscape, need to tackle with changes of  their means and ways for producing, transmitting, storing and consuming energy. A major transformation, which should radically modify the energy profile of our society, relates to the transition from an economy relying on fossil energy to more sustainable ones, based on renewable energy sources. Another  transformation relates to the potential diminution of relative contribution of nuclear energy gained traction in the wake of the Fukushima nuclear accident, which caused a number of countries to start rethinking of their energy policies, as for example Germany, where dismissals of nuclear plants is now being phased in. Within the EU, many states have nuclear power plants. Indeed, nuclear energy represents the fourth non renewable source of energy in the EU (36.7\% oil, 24.6\% gas, 17.7\% coal, 14.2\% nuclear), with France in particular producing  almost 80 percent of its electricity from nuclear energy sources.
The Fukushima nuclear power plant disaster has recently precipitated a call from the European Council for a review of risk and safety of all nuclear power plants, on the basis of a comprehensive and transparent assessment.  As a response to this call, 17 countries are participating, including 14 EU Member States operating nuclear power plants together with Switzerland and Ukraine. It is not clear whether the rest of the EU Member States will stop or continue with their nuclear power production. The future of nuclear energy production is certainly at a turning point.
\bigskip
\bigskip
\definecolor{shadecolor}{rgb}{0.941,1,0.941}
\begin{mdframed}[backgroundcolor=shadecolor]
\small
\begin{center}
\textbf{BOX 1:
Relevant Initiatives and Projects}\end{center}\par
\begin{itemize}
\item	``Towards Real Energy-efficient Network Design (TREND)"
The Network of Excellence on Energy Efficient Networking  \url{http://www.fp7-trend.eu/}.
\item ``On the Road to a Decarbonised Power Sector"  European Climate Foundation Power Perspectives 2030  (ECF, 2011) \url{http://www.roadmap2050.eu}.
\item ``A Roadmap for Moving to a Competitive Low Carbon Economy in 2050" 112 final (EC, 2011).
\item ``Proposal for a Regulation of the European Parliament and of the Council on Guidelines for Trans-European Energy Infrastructure" (EC, 2011).
\item ``Proposal for a Directive of the European Parliament and of the Council on Energy Efficiency" COM(2011) 370 final (EC, 2011)
\item ``Financing for a Zero-Carbon Power Sector in Europe": European Climate Foundation Roadmap 2050:  (ECF, 2011).
\item ``Enabling the low carbon economy in the information age" \url{http://www.theclimategroup.org}, SMART 2020 Report,(2008).
\item ``Energy infrastructure priorities for 2020 and beyond: A Blue-print for an integrated European energy network"	(EC 2010).
\item ``Mobilizing information and communication technologies to facilitate the transition to an energy-efficient, low-carbon economy", (EC 2009).
\item ``Probabilistic Forecast for 21$^{st}$ Century Climate Based on Uncertainties in Emissions (without Policy) and Climate Parameters" \emph{MIT Global Change Program Report}, Cambridge, USA, (2009).
\item``Energy Techology Perspectives, Scenarios \& Strategies to 2050", International Energy Agency Report (2010).
\normalsize
\end{itemize}
\end{mdframed}
\bigskip
\bigskip
\par
 Many services such as internet, communication and media systems, transportation, and urban infrastructures all depend on a reliable supply of energy, which is a key engine for social well-being and economic development. The security of energy supply is increasingly challenged by the fluctuating and uncertain nature of wind and solar energy and the simultaneous production of heat and electricity in combined heat and power systems. Currently, the energy mix, and in particular the share of renewable wind and solar energy, varies from country to country. Already today in countries with a large share of renewable energy the strain on the electrical grid is noticeable, creating problems unexpected until recently. In 2010, e.g., 127 GWh of wind power were wasted in Germany due to the lack of local power storage and insufficiency of the grid \cite{Ecofys}. These problems are expected to increase and extend to a global scale with increasing share of renewable energy and increasing complexity of future energy systems. In addition, the geographical distribution of traditional and renewable energy sources is going to emerge as a central problem from several viewpoints: technological and economic viability, environmental sustainability, and security of supply.
\par
The challenge consists in enabling local, national and supranational authorities to successfully address the energy system transformation including its multifaceted technical, business and social dimensions. This requires new methods relying, among others, on complexity theory and Information and Communication Technology as tools for interconnecting all actors, enabling a better control of the systems and adding new services throughout the entire value chain. New paradigms are needed, which will have to be based on:

 \begin{enumerate}
   \item multiscale cross-disciplinary vision,
   \item multilevel dynamical system approaches,
   \item adoption of global protocols encompassing: (i) fundamental phenomena underlying generation/dissipation of energy at the nano- and micro-scales; (ii)
     energy supply and demand at the meso-scales; (iii) overlaid by the internet and communication networks at the mega- and exa-scales.
 \end{enumerate}

 The paradigm shift in the future emerging energy global scenario will ultimately rely in the creation of a collective change in behaviour by influencing the individual awareness  to scale up to groups and, ultimately, to entire populations in order to achieve necessary changes in the profile of world energy consumption and management.
\section{The \emph{Emerging Energy Web}} The response to the energy shortage and global warming will require to understand and control the emergent global energy network increasingly fuelled by renewable sources. Such global energy network  could be as revolutionary in the coming decades as the World Wide Web did over the past decades. It will modify the way we use, transform and exchange energy \cite{Silbermann,Carreras}. More importantly, like the internet, it will affect our lifestyle, society and economy. Jeremy Rifkin, President of the Foundation on Economic Trends, Washington DC,  Maverick economist and adviser to German Chancellor Angela Merkel and other EU leaders, writes: ``We are on the cusp of a third industrial revolution ...The first industrial revolution brought together print and literacy with coal steam and rail. The second combined the telegraph and telephone with the internal combustion engine and oil. What we have now is the distributed energy revolution. We can all create our own energy, store it, and then distribute it to each other. Twenty five years from now millions of buildings will become power plants that will load renewable energy. We will load solar power from the sun, wind from turbines and even ocean waves on each coast. We can also make the power grid of the world smart and intelligent... Imagine the internet, only for energy.
Imagine that, as well as tens of millions of personal computers all linked together, exchanging information this way and that, you had tens of millions of personal power stations, pumping electricity to and fro. Imagine if, working together, they made fossil fuels redundant, resolved all our fears about energy security, and kickstarted a new era of peer-to-peer power sharing. Oh, and made a decisive impact on climate change" \cite{Rifkin}.
\par
\subsection{Renewable energy is potentially abundant}
Ultimately, all renewable energy on earth goes back to the radiation from the sun, which is the energetic driving force of wind, waves, rivers, and life on earth. Solar energy reaching the 113 million square kilometres cross section of planet earth amounts to  about 4 x 1024 Joule per year or 100 trillion tonnes oil equivalent. However, almost all of this energy is reflected back into the universe, partly after being transformed into heat, only a tiny part is converted into other forms of energy or used for other means. Indeed, not even one per mill is converted into biomass, and somewhat less of this amount into wind energy, wave energy being estimated at even less, namely about 3\% of the wind energy. Compared to these figures humanity's demand for energy is considerable but not necessarily excessive. At present, the annual human energy consumption amounts to about 17\% of the generated biomass or 22\% of the total wind energy; it is more than 7 times larger than global wave energy but only about 1/8000th of solar energy, corresponding roughly to one hour of global solar energy. Electricity consumption was about 12\% of the total energy consumption.
\par
These numbers indicate that humanity's ultimate source of sustainable energy can only be the sun. Indeed, if the same amount of solar energy were generated for human usage as is converted into biomass, every individual could consume 12kW, and humanity's energy problem would be solved; this figure corresponds to five times the average per capita energy consumption and more than the current per capita consumption of US citizens. In particular, less than 2\% of the solar energy available in the North Africa and Middle East deserts would be enough to cover today's global energy consumption. Accordingly the Desertec concept aims at promoting the generation of electricity in Northern Africa and in the Middle East by using solar power plants and wind parks, with the aim to transfer this electricity to Europe.
\par
This doesn't mean that other sources of renewable energy should be disregarded, quite the contrary: In the present situation where it is important to find a viable path towards global sustainable energy supply, all renewable energy sources that can make a contribution at competitive costs must be exploited. While many technological, economic, political and societal hurdles must be overcome before the ultimate long-term goal of a global solar energy supply can be reached, wind energy harvesting has just entered a new and rapid phase of development. Accessing wind energy from taller wind turbines, where the wind speed is stronger and more consistent, can make energy cheaper than coal. Equally, offshore siting of wind farms enables a greater energy transfer, although this is offset by the cost of establishing the difficult sites \cite{Marvel}. In certain coastal areas of the world, wind and water currents provide continuous waves over the ocean surface, transferring enough potential and kinetic energy to cover the energy demand of the local population. Suitably designed devices can now extract energy directly from ocean waves, their pressure fluctuations and vortices beneath. In the ultimate energy mix, wind, biomass, and, to a lesser extend, water and waves will still play a significant role, while fossil and nuclear energy will very presumably be limited to niche applications.

 Sources of renewable power vary considerably in different parts of the world, implying that our society should face problems of energy distribution rather than supply. Many detailed aspects related to these energetic problems can be found  for example in \cite{MacKay,Chu}, which give a balanced account of the current situation and future perspectives beyond all media hype.  Concrete cases the effort required to substitute non-renewable with renewable sources can be found in \cite{Marvel,Fagiano11,Franco,OECD08,JRC_Masera,ABB1,ABB2}.
 \par
\subsection{Electricity transmission is not running at its full potential.}
Key technology for the next major boost in the performance of power grids will come from power-electronic devices, network automation and monitoring and control applications. The combination of these layers - high-tech hardware, smart system-design, and intelligent algorithms - can leverage current and future transmission and distribution grids to their full potential.
\par
While some of these technologies can be added onto existing alternative current (AC) transmission and distribution networks, the full potential can be leveraged best while electricity transmission and distribution is slowly moved towards direct-current systems (DC). One significant improvement of AC lines is, e.g., possible with using flexible alternating current transmission systems, built from power-electronic devices, adjust physical parameters of the lines for maximum power (series compensation). They also absorb excess energy from disturbances and inject power into the system to stabilize the grid (shunt compensation). In some cases, system capacity can be more than doubled by these devices.
\par
Making full use of installed capacity, however, also includes the interconnection of neighboring grids enabling them to share reserves. To bridge long distances, cross water or connect asynchronous grids, high voltage direct current (HVDC) technology is required. In technical terms, HVDC technology supports load flow, voltage and power oscillations control, reactive power support, flicker compensation, voltage quality, handling of asymmetrical loads and handling of volatile loads. Very-high-voltage solutions are primarily focused on long-distance, point-to-point bulk power transmission. A typical application can be the transmission of thousands of megawatts from remote hydro sources to load centers. Additionally, there exists ``lighter" HVDC technology with which integrating dispersed, renewable generation, e.g. wind power into existing grids is made possible and economical. This lighter solution It is also used for smart transmission and smart grids due to its great flexibility and adaptability. The first commercial HVDC installation was built in Sweden in 1954, there are now over 120 HVDC systems in operation in the world.
\par
Regarding network automation, already now, many solutions exist to facilitate distributed generation, participation of the electricity user in liberalized markets and an increased use of automation (including operational distribution automation, active demand management and automatic meter reading). A major step further step is foreseen with the increased availability of network load data and algorithms that allow for a much more predictive and interactive operation of the distribution systems.
\par
Such smart distribution grids of the 21$^{st}$ century will require innovative operation centers where grid operators will have a comprehensive view of the distribution system, including system status and monitoring, control, outage response, planned work, optimal equipment loading, and improved control over distributed generation, energy storage and demand response resources. These integrated distribution operations centers will help distribution companies in their mission to meet the goals of customers, owners, employees and society itself.
\par
Still the problem of storage must be addressed. Even with the pervasive use of ICT-based smart-grid network technology it will not be possible to completely balance supply and demand. In particular, in an electrical power system largely based on fluctuating renewal energy, CO$_2$-neutral power production will stay behind its possible level without sufficient storage. As the electric grid itself has no storage capacity, adequate storage technologies are required on different scales, from local grid stabilization to seasonal storage, to balance residual local and temporal mismatch of production and consumption. Only few technologies have the potential to contribute on a wide scale of power (kW - MW) and energy (MWh - GWh) \cite{A2}. Even e.g. Switzerland is very unlikely to meet its target of considerably increasing the share of renewable energy by mainly relying on pumped hydro storage, in spite of its particularly high fraction of pumped hydro storage plants. Indeed, even after completing the current extensions to 4.2GW total \cite{A3} compared to the average load of 10GW, energy can be accommodated only on the scale of days. Therefore, the grid of the future is in dire need for solutions that offer dispatchable loads of considerable size and means of transferring energy over long periods of time.
\par
Since all traditional storage technologies are limited either by capacity or by local constraints, solutions beyond the traditional ones should be considered. In contrast to the electric grid, the European natural gas grid has an existing storage capacity of the order of 650-860 TWh \cite{A4}. It is therefore intriguing to use this already present storage capacity to store surplus electric power after conversion to hydrogen and synthetic natural gas through hydrolysis or methanation, to transport it to the location needed and using it for powering cars or reconverting it to electrical power in CHP-blocks. From a network perspective this ``power2gas" technology means coupling the presently separated electrical and gas grid. This adds a new dimension to the already complex tasks ahead, but at the same time has the potential for considerable relief.

\subsection{Energy efficiency - the low-hanging fruit}

Despite of the huge potential of smart grids, it should not be forgotten that still many millions of tons of fossil fuels are burned each year to generate electricity. Through inefficiencies, from the gathering of these energy sources to their eventual consumption, needless amounts of carbon dioxide are contributing to global warming.
\par
Faced with the urgency of mitigating climate change, our best hope of significantly reducing emissions is to use energy more efficiently. Estimates suggest energy efficiency improvements could deliver half the cuts in emissions needed to slow global warming over the next 25 years and by using energy more efficiently these precious resources will last longer and will also save money.
\par
Variations in energy efficiency across the world give a sense of what can be achieved with todays technologies. The most efficient economies generate almost 16 times more GDP with the same amount of energy than the least efficient ones.

\par
\begin{figure}
\center
% Use the relevant command for your figure-insertion program
% to insert the figure file.
% For example, with the option graphics use
\resizebox{0.5\columnwidth}{!}{%
\includegraphics{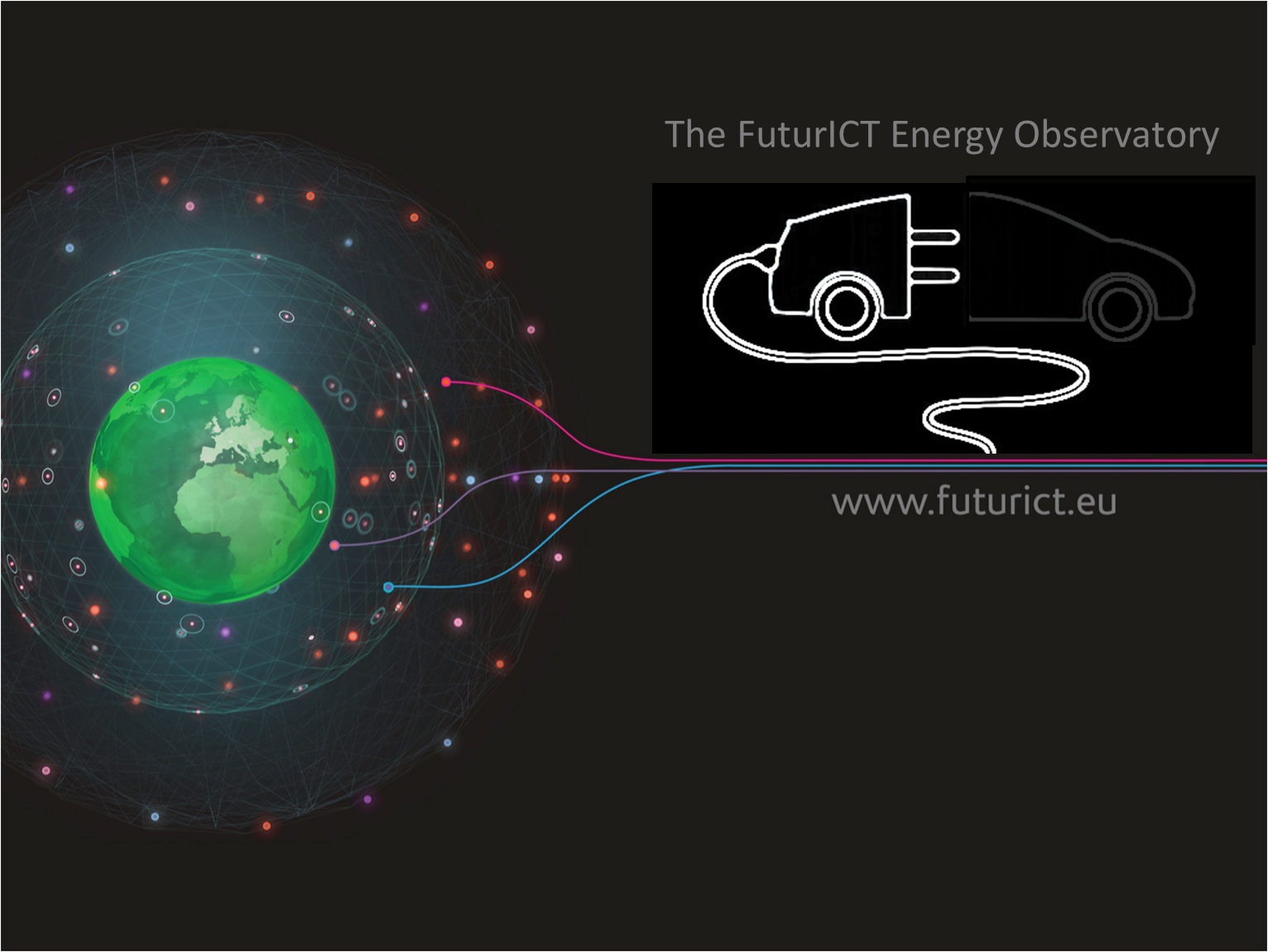}}
\caption{The Energy Exploratory will focus on the global energy web as a component of the FuturICT multilayered network structure (\url{www.FuturICT.eu}).  }
\label{fig:plugin}       % Give a unique label
\end{figure}

Altogether the challenge consists on the creation and management of a global energy network, which will be affected by a variety of new issues arising from the political and social impact of this upcoming global transformation. The future energy scenario will mostly be concerned with solving problems related to supply security of the future energy network. This implies innovate solutions for energy distribution heavily relying on ICT, and including the coupling of hitherto separate networks. This new way of securing and managing energy will also  need another paradigm shift in societal organization. Understanding the underlying dynamics of these transformations will certainly contribute to meeting the grand energy challenges faced by the global community. These issues will be central to the FuturICT Energy Observatory (Fig. \ref{fig:plugin}).

\section{Single ants  are not smart, but their colonies are.} How concepts derived from the study of swarm intelligence \cite{Sunstein,Surowiecki,Buhl,Couzin,Helbing92,Cavagna,Gregoire} can help people to understand and manage the emergent global energy system?
\par
The possibility to learn and borrow organizational schemes from other scientific communities should be considered as one of the main paths to achieve this goal.
\par
 Just as an example consider fish swimming in a swarm, which use the vibrations induced by the vortices generated by the other fish around them to swim in the most efficient way. If one measures the speed and mass of the fish, to make an estimate of the kinetic energy and compare it with their muscular energy, one can notice that fish swim faster than their muscular strength would allow. This energy gain is possible because the vibrations induced by the vortices generated by the fish swimming in front of them and by gliding between them allow fish to use less muscular energy and, at the same time, propel them at a much faster rate \cite{Alexander,Williams,Fish,Weihs,Svendsen,Weimerskirch,Wootton,Allesina}.
\par
It is not surprising that the behaviour of people or animal groups such as fish schools, bird flocks or insect swarms has been associated with the concepts of ``collective order". The same is true for people in a crowd.   The emergence of this collectively ordered behaviour seems to be generated by the repeated local interactions among the individuals as in the vortex induced vibration of the fish example. There are many interesting features of the dynamics underlying such phenomena that might be  worthy of deeper investigation in the framework of the FuturICT energy Exploratory endeavour. The exact nature of the underlying ordering and coordination mechanism is not straightforward and depends on each community. In the fish clustering example described above,  it might be that an individual can take advantage of the synchronization with the movements generated by others to reduce the energy cost of the locomotion. This elaborate version of slipstreaming enables the swarm as a whole to propel at a velocity that comes close to what could be achieved with the aggregated muscle strength of the fish if the warm were a single entity.
  \par
 The universal feature of this highly integrated collective behavior can more generally be related to the way the information is acquired, transmitted and processed by the individuals within the group. Alignment among individuals can allow the transmission of information about an incurring change in direction. Amplifying local fluctuations through positive feedback is important  when threats, are detected, may it be a predator by prey animals, technological disruption by people, or more general, social turmoil by strongly interacting individuals in a complex system.
 \par Fish schools,  insect swarms, people crowds  move and interact by relying on the information received through the aggregated group by deploying an effective sensory range  wider than the one that would be available  by relying only on the individual perception. This increased sensitivity can enable individuals in a crowd to avoid the time delay and energy cost  associated with the integration over time  and space of the information. One interesting aspect of the nature of these local interactions is how they  enable the behavioral control to be uniformly distributed, as opposed to a hierarchical top-down control. Such distributed and participated coordination depends on the entire group rather than just on few individuals implying that the coordinated activities of the groups are intrinsically more resilient. The resilience is enabled by  the encoding of the information across a wide range of spatial and temporal scales, which involve the existence of multiple collective stable states and ultimately imply further resilience. Such collective behavior has in some cases been referred to as ``the wisdom of crowds" \cite{Surowiecki}.
\par
 The deployment of a ``wisdom of crowds'' approach  in the context of the global energy systems is one of the challenging aspects of the future  system management. Studies regarding how agents do react to local voltage variations to modulate energy consumption and supply in case of local generation might help to shed light on these complex dynamics. The adoption of schemes from different communities and organizations and their adaptation to innovative socio-technological system areas should open unthought-of scenarios and pave the way to deeply innovative changes to challenge  the current organizational and institutional policies concerning the management of  energy. Decision-making by individuals  in terms of energy related problems should become intimately coordinated over a wide set of temporal and spatial scales within groups and large aggregates as in the swarm intelligence examples. Furthermore, optimal energy distribution schemes across complex networks at the megascales should be scaled up from the behaviour of complex individual networks at the meso- and nano- scales.
Concepts deriving from modeling and simulation approaches developed in the framework of Complex Systems science are envisaged to be vital to unravel new aspects of collective behavior, intrinsic resilience, and avoidance of systemic instabilities of the new energy web.
A full and effective coordination of agents can be further facilitated  by implementing the most appropriate technological changes in combination  with the creation of social, financial and reputation incentives. New technology and new market rules are needed to create new business models and make every innovative solution  economically attractive. To simulate the market  behavior  upfront will be crucial  to know what  happens before we implement in the future energy scenario.
\begin{figure}
\center
\resizebox{0.5\columnwidth}{!}{%
\includegraphics{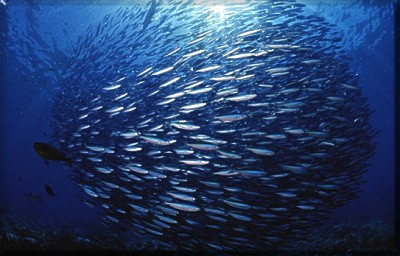}}
\caption{Fish use vortex induced vibration generated by the fish swimming in front of them and by gliding between them to swim in the most efficient way from an energetic point of view.}
\label{fig:phone}       % Give a unique label
\end{figure}
\par
\section{Single power plants are not smart, but the \emph{Emerging Energy Web} will be.} The introduction of smart-meters and the development of the smart-grid is a true revolution in the electricity industry \cite{Beyea,Schlapfer}. It changes  patterns of consumption, and the business models of generators, distributors and retailers. Due to smart meters, power generation becomes a decentralized network of digital systems.
Smart grid is a combination of electronics, information technology, management and reporting software. In this context, the implementation of new business models, which help to improve industrial efficiency and to reduce carbon dioxide emissions should be facilitated. From a retailers' perspective, the smart-grid represents both a potential opportunity for reducing  fixed costs and improving customer service, and a threat of increased competition. There is an urgent need to  develop computational models for different market designs, to address, e.g., how the  equilibrium condition depends on market rules  and what the properties of this equilibrium are. This understanding is essential since market failure due, e.g., to collusive behavior or instability of infrastructure as well as the resulting  overall consumers benefit, must be correctly understood, before a  particular market architecture is chosen for implementation. Developing and analyzing reliable models for competitive market architecture and taking into account grid restrictions, is also essential before any practical implementation.
\par
The social dynamics of adoption of smart energy production systems and energy efficiency schemes should  also be taken into account. Being parsimonious comes at a cost, and it is not easy to discover whether the developed countries are ready to bear this cost or they prefer to trade short term advantages for long-term suffering. Complex social sciences systems need to be modeled, in order to be able to find the right spots to ``nudge" people into buying in the attitude changes required by the future energy system.
The accomplishment of such challenges will heavily rely on the other components operating within the FuturICT platform \cite{N1}.  The ``Planetary Nervous System" will allow one to access the ubiquity of sensors for energy production, transportation, storage, and consumption at the different scales \cite{N3}. Furthermore, it will allow one to access  behavioural and economic patterns generated by the energy user profile. If enough data will become available and appropriate models are developed, simulations and probabilistic predictions can be attempted \cite{N2e6}. The ``Living Earth Simulator" will  allow the evaluation of the impact of decisions - as concerning policies, price schemes, new technology deployment- without having to implement them.
New energy policies and technologies will be more easily accepted by citizens and stakeholders by ensuring a deep and continuous involvement in the decision processes. This will be made possible by the interfaces provided by the ``Global Participatory Platform" \cite{N4e5}  and will represent the key to the success of the Exploratory.
\par
Important scientific and societal issues  of the upcoming energy scenario, are for example:
\begin{itemize}
\item  How fundamental phenomena at different scales are related and affect each other?
\item  What technological advances  should be envisaged to accelerate and make efficient such integrated vision of the energy system?
\item  How the multi-dimensional multi-actor energy system will ensure the economic, environmental, security, and social requirements as a whole?
\item  How the future global energy scenario will  be affected by different threats and how can one help in managing these risks?
\item  How the foreseeable transformation of the energy system will have an impact on society and might be themselves affected by crime, terrorism, international politics, environmental and economical changes?

\end{itemize}
These questions need to be addressed through a multifaceted approach founded on a cross-disciplinary perspective. In the remainder of this manuscript, the above themes are developed and discussed in deeper detail and some suggestions to the way forward are given.
\section{Geopolitical Challenges and Relevant Initiatives}
\label{GEOPOLITICAL}

The European Union as world wide largest energy importer and second largest energy consumer faces an extra problem in the global energy scenario. Nearly half of its total energy demand is supplied from outside of non-member states like Russia, the Middle East, North Africa, and the Caspian Basin. For European countries, reliable energy supply is entirely based on good relations with neighboring countries in North Africa, the Middle East, Russia, and Asia  \cite{Corridors}.

The Desertec initiative mentioned above will represent a major advance in the transition towards renewable energy sources as it addresses the problem of transferring  plentiful solar energy from the Middle East and North Africa for distribution in Europe \cite{DESERTEC}. The need of transmitting electrical power from one region to another is not new. European grid allows countries like Denmark to boast about its relatively large fraction of electricity provided by wind, without the cost of in-house storage. When Sweden is lacking wind or hydroelectric capacity, excess generation capacity elsewhere can supply what is missing. However, Europe aims higher, and Desertec is a centerpiece of its plans to dramatically increase renewable sources in its energy supply mix. Although Desertec will stretch over only 15\% of the Northern Hemispheres latitudes, it will be a first step toward Buckminster Fuller dream of a worldwide electricity grid to allow uninterrupted solar and wind electrical power generation, with the grid serving in lieu of local storage.
\begin{figure}
\center
\resizebox{0.5\columnwidth}{!}{%
\includegraphics{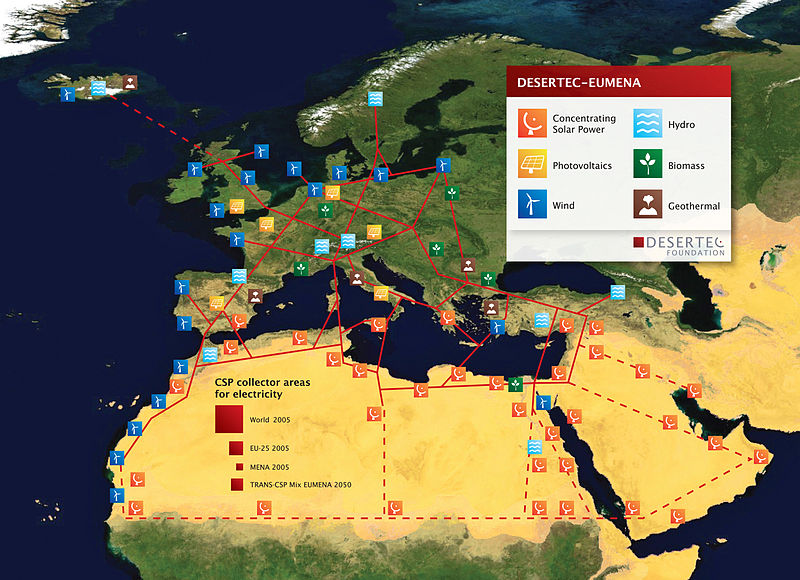}}
\caption{DESERTEC EU-MENA Map: Sketch of possible infrastructure for a sustainable supply of power to Europe, the Middle East and North Africa.
The red dashed squares represent the total surfaces needed for solar collectors plants to cover the current electricity demands
for the world \cite{DESERTEC}.}
\label{fig:phone}       % Give a unique label
\end{figure}
Desertec is both more ambitious and more complicated than similar projects. The reason is politics and political instability, especially in light of the recent developments in the MENA region. The present European grid succeeded in an environment of political stability after the second World War. Can Desertec succeed in the new political climate that has enforced regime change in North Africa? While, without question, countries like Morocco and Egypt have very serious interest in the program, involving the latter may well imply undersea connections with materials and other challenges, unless neighboring countries participate. Morocco, already connected through Spain, is presently in a better situation. Coupling the electrical grid and the gas grid may complement this at a future stage. Nevertheless, looking at the European grid today shows how fast the interconnections can grow once the basic ingredients are assembled. This should provide, in the face of any pessimism, further impetus to push the project forward. Indeed, it is possible that the materials and other research developments needed to make Desertec possible (e.g., self-cleaning panels, increased thermal storage, tamper-proof systems, safe and sealed cables with appropriate sensor systems) may find first use elsewhere or in a very small part of the proposed Desertec area. When these developments are in place and tested, and the general political climate are right, the plan can be quickly realized.
\par
Medgrid is promoting new high capacity electricity links with the main motto ``No Transition without transmission", underlining that the  transition to a decarbonized energy future cannot be achieved without an improved and strengthened grid, in Europe and the neighboring countries \cite{MEDGRID}. Desertec and Medgrid are complementary and mutually reinforcing, the first focusing on energy generation and the second on energy transmission: an increased collaboration between these projects will create even stronger synergies in the international energy scenario and burst similar joint initiatives in other souther-northern close regions of world. The third pillar should be a powers2gas project exploiting the huge storage capacity of the gas grid, but to our knowledge no European project in this direction is under way.
\par
The above initiatives, though differ for their specific technological targets, share a certain number of risks and uncertainties since all rely on the connectivity of Europe to neighbouring countries.
The Energy Corridor Optimisation for European markets of gas, electricity and hydrogen (Encouraged) project was launched in 2005 to analyse the performances of the corridors \cite{ENCOURAGED} and in particular:
\begin{description}
\item[(i)]   Assess the economic optimal energy (electricity, gas and hydrogen) corridors and related network infrastructure for connecting the EU with its neighbouring countries and regions.
\item[(ii)]  Identify, quantify and evaluate the barriers to and potential benefits of building optimal energy corridors connecting the EU with its neighbors.
\item[(iii)] Propose necessary policy measures to implement the recommended energy corridors with a focus on investment and the geopolitical framework.
\item[(iv)]  Organise stakeholder workshops and seminars to discuss the results and findings and reach consensus among scientists, stakeholders and non-governmental organizations and validate project results.
\end{description}
Investments in such huge infrastructure projects are certainly subjected to different types of risk, as:

 \begin{itemize}
 \item Market Risk: uncertainty on price and volume
 \item Regulatory Risk: impact of market rules and regulation
 \item Political Risk: uncertainty relating to international relations and often involvement of several transit regimes  (USA)
 \item Technological Innovation Risk
\end{itemize}
These risks create shadows on the profit horizons and therefore prevent the big investors commitment. Therefore, to develop corridors for both renewable and traditional power supplies, the backbone of the emergent energy web, one important action should be particularly addressed on managing and reducing the risks mentioned above through a fair sharing of resources \cite{Jakob,Carvalho12,Corridors}.
\begin{figure}
\center
\resizebox{0.5\columnwidth}{!}{%
\includegraphics{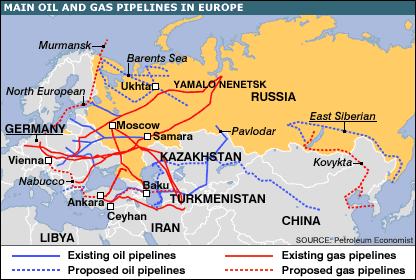}}
\caption{Scheme of the main pipelines bringing oil and gas to Europe.}
\label{fig:phone}       % Give a unique label
\end{figure}

\section{A multiscale walk down the \emph{Emerging Energy Web}}
The Divine Comedy can certainly be considered the most ambitious and famous multiscale walk that human mind might have conceived. It might be instructive trying to figure out what kind of effort should have led to imagine such a complex structure of up-and-down circles  of varying size, depth and people occupancy from  hell to paradise.
\par
We need today a comparable effort to figure out and comprehend the multiscaled complexity of our strongly connected world.
As noted above, any attempt to tackle the  emerging future energy challenges would require a multiscale  and multidimensional path. The increasing size and density of communication infrastructures with the diversification of interconnected people and objects is creating a hierarchical structure of energy sources and sinks ranging from exascale data management to i-phone powering.
\par
Shekhar Borkar, Director of Extreme-Scale Technologies at INTEL, replied to a question on the possible solutions to the Exascale Data Power Challenges:
``If engineers can use new technology to create an exascale system that consumes only 20 MW of power, the same technology can also be used to dramatically lower the power consumption of lower performance systems, to the point where giga-scale systems consuming only 20 milliwatts of power can be used in small toys and mega-scale systems that consume only 20 microwatts could be used in heart monitors". Shekhar Borkar pointed to the energetic relevance of ``local computing" (nanoscales
) at a time when everyone is talking up the virtues of ``cloud computing" (exascales and up) and that communications technologies used for moving data, including Bluetooth, Ethernet and Wi-Fi, use far more power than those used for local computing within a chip or system.
\par
Overall, the true challenge for the near future is to reason by scaling-up and down energy problem solution, according to the approach envisaged by the FuturICT Exploratory on Energy, where expertise from Complexity Science will be joined to those of Computer Sciences to achieve as broad as possible a comprehension of the \emph{Emerging Energy Web}.
\par
In the following subsections, we will attempt a multiscale walk discussing a few examples of systems and problems starting from the exascales and ending to the nanoscales of the envisaged Energy Web.\\

``\emph{But follow, now, as I would fain go on, ... }'' (Dante, Canto XI, Inferno)

\subsection{From Exascales to Megascales}
\label{Scientific}
This is the realm of Internet, Satellite Communications, Cloud computing and, generally, of any Communication and Data Network.
Research in the area of energy-efficient communications and networking has a long tradition, and has been initially focused on energy-efficiency for battery-driven devices, such as satellites, portable computers, mobile phones and sensors. Up to a few years ago, very little concerns were addressed on energy-efficiency in non-battery-driven communication systems and networks. The general attitude was to consider the power necessary to run networks be granted: networking equipment (such as wireless access points, ADSL lines, racks of switches, routers, data centers, etc.), and often user terminals, were assumed to be powered on 24/7, even if lightly used, or not used at all (e.g., in off-peak traffic conditions, or for fault protection).  Today, this attitude is urged to change for the reasons mentioned in the Introduction.  In the developed world, global warming  has raised a general concern about energy consumption. Electric energy has become an expensive resource. Some European wireless network operators declare that their energy costs are double than their operating expenses. The exponentially increasing request of network access must cope with the limited power availability, so that energy-parsimonious approaches are becoming a must.
\par
The academic research community has quickly recognized the importance of the energy-efficient networking issues. Researchers are now exploring the potential of green communications, including wired and wireless access, backhaul and core networks, home and enterprise networks. A number of symposia, workshops and panels were recently organized specifically on the issue of energy-efficient communications  \cite{Yick,Yu,Conte,Ajmone,Zhang,Koutroumpis}.
\par
The interest of industrial research in the energy-efficient communication area has been extremely keen since the very beginning. The growing of the industrial interest in energy-efficient networking culminated this year, when Alcatel-Lucent Bell Labs announced the launch of \emph{Greentouch} \cite{greentouch}, a global consortium whose goal is to create the technologies needed to make communications networks 1000 times more energy efficient than they are today. As the press announcement observed thousand-fold reduction is roughly equivalent to being able to power the worlds communications networks, including the Internet, for three years using the same amount of energy that it currently takes to run them for a single day.  The common target to the academies and companies participating in the consortium being to invent and deliver radical new approaches to energy efficiency that will be at the heart of sustainable networks in the decades to come. The goal of a 1000-fold increase in energy efficiency appears extremely ambitious, especially in the medium-short term, but it is based on models and simulations derived from an analysis of the fundamental network properties, specific technologies (optical, wireless, electronics, processing, routing, and architecture) and their physical limits which envisaged the potential for the networks to be 10,000 times more efficient than today.
\par
The overall focus would be on an adaptive energy efficient network design which will require a dynamic reduction of the global energy network consumption by modulating the network capacity to the actual data traffic, while guaranteeing the optimal performances of service. Traditional telecommunication networks have been designed and planned to guarantee a given service quality under the peak hour traffic demand, implying the installation, deployment and powering of an exceedingly large number of devices that are by design underutilized during off-peak hours. By suitably turning off and on these devices, it becomes possible to adapt network capacity to the present need, reducing the waste of energy. This intuition has been studied in the context of UMTS access networks, WiFi access in large enterprises, Internet backbone networks, and optical networks. A theoretical study based on random graph theory has also been pursued to assess the global energy saving that the previous policies can possibly achieve when applied on a large scale. Comparative study between optical and electronic technologies are trying to see what could be the benefit of adopting optical transport technologies instead of electronic transmission and switching equipments in terms of energy waste reduction.
\par
With the exponential increases in the amount of channeled information, the contribution of the ICT industry to the global carbon footprint is forecast to double over the next ten years. While purely technological approaches to reducing energy consumption are necessary and welcome, they will need to be complemented by viable solutions in the social, economical, legal and regulatory fields, as well as in the energy generation and transport sector. Interdisciplinary, holistic, ambitious and visionary solutions are required, to achieve a lasting impact in the long term.
Cloud computing is an innovative networking technology exploiting the remote handling of security and lighting or even the settings on appliances and other energy-consuming devices. Energy management will certainly see an increasing trend for cloud-based services and systems going forward.
Growth in cloud computing might have important consequences for both greenhouse gas emissions and sustainability. Thanks to massive investments in new data center technologies, computing clouds in general and public clouds in particular are able to achieve industry-leading rates of efficiency. Simply put, clouds are better utilized and less expensive to operate than traditional data centers. As a result, we anticipate that much of the work done today in internal data centers will be outsourced to the cloud, resulting in significant reductions in energy consumption, associated energy expenses, and greenhouse gas emissions from data center operations versus a business as usual scenario.
\par
Several projects on energy-efficient networking are now funded, both at the national and international levels. The goal is to quantitatively assess the energy demand of current and future telecom infrastructures, and to design energy-efficient, scalable and sustainable future networks, considering both the backbone, and the wireless and wired access segments.
Funding research in energy-efficient networking and computation is very rewarding, but does not cover the whole set of issues raised by the present situation.  Most research projects launched up to now have tackled the problem along only one dimension: Technological Advance. However, networking, as the whole telecommunications sector, is much more complex than just technology.  Decisions are often based on economic considerations and on social impact, rather than on technical quality. A multidisciplinary approach should be pursued to achieve really significant results, which can seriously impact the amount of energy consumed by telecommunication systems and networks. Indeed, as we found out in some of our technology-oriented research, once a solution optimal from the technological standpoint is found, several roadblocks remain, related to energy generation and transportation, to social, economical, legal and regulatory issues. This is the urgency and the added value of considering this issue in an open  multidisciplinary research platform like FuturICT.

\subsection{From Megascales to Mesoscales}
\label{MESOSCALES}

In the multidimensional multiscale energy scenario, the intermediate  scales are mostly inherent to the energy Supply-Demand, thus are directly linked to and affected by the   perception and interactions humans have with them. Importantly the mesoscopic scales are strongly interrelated to the market and geopolitical issues.
As stated in the Introduction, our society will face major challenges over the next decade mostly related to the transition:
\begin{itemize}
\item  from an economy relying on fossil energy to a low-carbon economy, based on renewable energy sources, and
\item   triggered by the recent Fukushima accident in Japan which have caused a number of EU countries to rethink of their energetic policies mostly based on nuclear energy (Germany, Switzerland)  and  are evolving towards different energy scenarios.
\end{itemize}

A low-carbon economy would have a much greater need for renewable sources of energy, large capacity energy storage facilities, energy-efficient building materials, hybrid and electric cars, ``smart grid" equipment, low-carbon power generation, and carbon capture and storage technologies. These needs can only be accomplished if we acquire the necessary tools to solve a multi-faceted supra-national problem in terms of technological advances, economic strategies, political  and  social preparedness. Economic and political commitment will probably be achieved because, throughout history, the nations that have ruled the world, economically, politically, and militarily are those who have innovated to meet social needs.
\par
Up to now, electricity generation has relied mainly on fossil fuels such as coal, which is a major source of CO$_2$ emissions. While the electricity sector has reduced emissions by increasing natural gas and nuclear generation and promoting energy efficiency and conservation, these reductions will remain limited without the concurrent development of renewable energy sources such as solar, wind and hydro-power. Finally, the rapid expansion of renewable energy can be technically difficult to integrate into the national grid systems and also economically costly. A case in point is wind generation: often located far from population centers, it requires expensive transmission and system integration even if plentiful. One central aspect concerns the effectiveness and efficiency of existing policies in promoting renewable energy including feed-in tariffs, solar energy programs, wind energy development, and renewable energy support.
\par
The economic and political aspects of the period of transition for energy policies need to be carefully assessed. These aspects include electricity market price and zonal market price behavior, efficient electricity pricing and investment, regulatory behavior, policy interaction and redundancy, the long-term impact of hydrogen fuel and synthetic natural gas development" on electricity price and investment, bio-fuel market behavior, renewable fuel project risk, benefits of rural solar home systems, the effect of plug-in electric vehicles on electricity costs, achieving a low-carbon electricity system in China, and the institutions and processes necessary to scale up renewable energy emissions. At a global level, the advantages of renewable energy are largely undisputed: renewable sources of energy emit little or no GHG emissions and thus represent a key mitigation action for combating climate change. However, at a local level, the environmental attractions of renewable energy are less clear due to a range of impacts including visual intrusion, noise, land and water alienation, ecosystem disturbance and in some cases local pollution of air, water and land.
\par
Climate warming  is by far the most important challenge humanity has faced in our history; the future of our society as we know it depends on solving this problem and developing a global collective response to the related issues.  In this sense many industries, individuals, institutions, and corporations have adhered to ``Climate Capitalism" as a way to address the climate warming by addressing the actual economic crisis. However, the money and commitment of capitalism is not enough if the technology and the workforce are not up to date. Technological change  can be accelerated through increased invention and through economic incentives, but the need for a sufficient and qualified workforce ultimately drives the success.
\begin{figure}
\center
% Use the relevant command for your figure-insertion program
% to insert the figure file.
% For example, with the option graphics use
\resizebox{0.5\columnwidth}{!}{%
\includegraphics{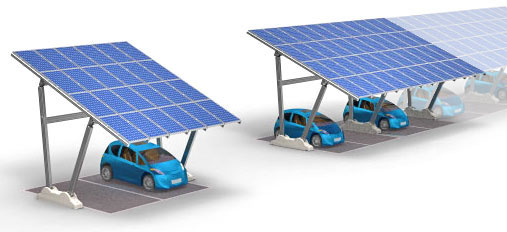}}
\caption{Completely new user-and-energy- friendly solutions should be envisaged, put in place and transferred to the market.}
\label{fig:phone}       % Give a unique label
\end{figure}
\noindent
In their plans to move beyond heavy dependence on fossil fuel imports  or to maximize export revenues from domestic oil and gas reserves, the North African states stand at a crossroads in terms of energy policy: they are interested in adopting renewable energy and/or nuclear energy, which presents opportunities for a strategic realignment of national development paths. Placed in the global sunbelt, rich in wind, geothermal and hydro power resources, the North African countries boast abundant potential for renewable energy production.
\par
Although a series of clean energy policy initiatives have recently been introduced in the region, renewable energy resources largely remain untapped. Current efforts to establish large-scale solar power exports to the EU including the Mediterranean Solar Plan and the Desertec industry initiative, anticipate a substantial uptake of renewable energy in North Africa, but so far there has been only limited buy-in by Arab political regimes, although recent political events in the region may presage new energy policies and initiatives. At the same time, several North African states are currently seeking to obtain civil nuclear power for several reasons: to meet rapidly growing domestic energy demand, to protect exports revenues from fossil fuels, to demonstrate national technological advancement, and to increase geopolitical recognition in the face of the growing regional influence of Iran and its nuclear programme. Across North Africa, there is extensive uranium prospecting and exploration being undertaken mainly by Australian, European and Russian companies.
\par
A key challenge for the 21$^{st}$ century is to change both producer and consumer behavior to realise efficiency gains in energy production and consumption and load management in view of fluctuating production.   This is the essence of the smart grid and its potential benefits to the world for its future well-being.  Essentially, it requires us to develop insight into the behaviour of how energy technology, ICT, and social structures can come together and interact in a mutually desirable way for the benefit of our ecosystem.
\par
If, by means of ICT infrastructure, a change of attitude could be achieved,  which enables a successful synthesis of central and consumer controls of their energy usage, we could have:
\begin{enumerate}
\item	more even and efficient use of the renewable power production capacity together with higher supply security.
\item	a reduction of the  carbon footprint for Europe.
\item	a reduction of daily-peak  demand for electricity and thereby a reduced demand for larger cable capacity to deal with peak demand. Such substantial changes in infrastructure are extremely expensive at a time of rapid world increasing costs of metal commodities.
\item		a further move towards the internationalization of electricity demand and supply, and attention to the increased need for electricity flow at a much greater level across inter-national boundaries.
\end{enumerate}
The key drivers for these development are at the global scale the issue of global warming and the increasing emission of carbon dioxide. To adapt to new energy sources which are relatively environmental friendly at both the national and international scale - requires the further development and implementation of  renewable energies; specificallywind and (to a lesser extend) tidal energy for the short term, and solar energy for the longer term.
\par
In parallel with increased production from renewable sources is the need to improve power transfer capacity across national borders and storage capacity, to deal with the diverse locational and temporal issues of wind power production  and distant storage facilities. Within Europe this means transfer of power from the wind rich north to the storage rich south in the short to medium term and as the century progresses to the problem of transfer of solar energy from more remote areas such as the Sahara Desert and North Africa. At this horizon one could expect the power2gas technology to be operational, which will be a relief of the electrical power transfer.
\par
There are major challenges in addressing the problems of infrastructure capacity \cite{Arrowsmith10}.  The complexity and inaccessibility of the infrastructure network makes it extremely costly to upgrade. Essentially, putting more copper cables under the roads and streets of Europe is currently seen to be prohibitively expensive and not a first option.  Fortunately, the danger of overloading the network is a rare event.  For example, the risk of under capacity, with the concomitant chance of an escalating blackout, is currently a potential hazard for a total of  a few days per year in the UK.
\par
One way of decreasing the risk of power line overload and the associated blackouts is to flatten consumer demand by reducing the peak consumptions, which typically occurs twice a day, in the morning and the early evening.  To deal with the daily  danger of power grid under capacity is a key problem, but nevertheless a predictable one. There are other, singular and unpredictable, events (e.g. in the wake of natural catastrophes) which may result in network disruption.
 The key requirement of all national grids is to maintain a careful matching of supply and demand to ensure that the grid does not desynchronize from its set-point frequency.  Deviation of the frequency as a result of mismatch  can  precipitate  major blackouts of supply at both national and international  level  if the network cannot balance the mismatch quickly enough.  There are well documented cases of pan-European and pan-American blackouts.  The major European blackout of November 2006 led to the missive from the Energy Commissioner  Andris Piebalgs (\url{http://www.continuitycentral.com/news03043.htm}) that ``Europe should draw lessons from this blackout that left millions of Europeans in various Member States without electricity and develop stronger network security standards".
\par
The need for the efficient and shaped use of electricity by the consumer is a key component of the smart energy agenda to help in achieving an even more reliable infrastructure. This is particularly needed as we move towards a scenario of renewable energy, which is currently mainly sourced by wind-power which induces volatility and uncertainty into the supply chain of power.
\par
The precise way in which the supply-demand structure can be controlled and programmed to effect the required smoothing of daily energy profiles is a major problem already recognized by governments. The UK Audit Office, HC109, 30 June 2011: Preparations for Roll-out of smart meters observes:
\begin{itemize}
\item	There are major risks that the Department of Energy and Climate Change must address to achieve value for money from its £11.3 billion national programme to install smart electricity and gas meters in all homes and smaller non-domestic premises in Great Britain from 2014 to 2019.
\item	The Department assessment that consumers with smart meters will annually use 2.8\%  less electricity and 2\% less gas than consumers who do not have smart meters is based on estimates contained in a 2008 review of trials and international experiences, but is also informed by more recent reviews.
\item  It estimates the programme will deliver efficiency savings to energy suppliers; and en-able energy consumers to change and reduce their energy use, resulting in savings on their bills and environmental benefits. There is, however, uncertainty over how much, and for how long, consumers will change their energy use and therefore whether the benefits will be fully realised.
\end{itemize}
Smart appliances are devices with sensor and remote connectivity that can be easily controlled at consumption peaks of demand and could be a key way of enabling current infrastructure to have an increased useful lifetime. It is important to note that a random element in the response of consumers is crucial.  The grid requires the avoidance of massive and fast swings in consumption. The consequences would be as catastrophic if all consumers switched off their appliances as if they switched them on: the discontinuity between supply and demand would result in major blackouts.
\par \url{SmartgridNews.com} reports that Household appliances (heating and cooling systems, refrigerators, electronics, hair dryers) account for 60 to 90 percent of the residential electricity consumption in the U.S., depending on whose reports you read. More and more of those appliances are becoming ``grid-aware" and gaining the ability to monitor and report their own usage and to increase or decrease their electricity usage by remote command.  Smart appliances are an extremely useful way of making well-controlled adjustments to consumption as they can be randomly triggered to switch off in appropriate quantities.
\par
There is also substantial scope, and indeed need, for a smart implementation of the increasingly ubiquitous ``charging device" for personal and household battery powered devices.
Nevertheless, it is clear from the National Audit Office that getting consumers to work with these potential energy savers and shapers is a major challenge.  Thus the  grid/consumer interaction is critical to the success of the new initiatives, which could address many major problems of energy usage.
\par
Demand side management issues are considered in \cite{Strbac}. Establishing a business case for the application of these enabling technologies in such a complex environment is not straightforward. Clearly, no individual recipient of the services (e.g. producers or distribution network operators) is interested in maximizing the overall system benefits that can be achieved by trading off the benefits between individual segments of the industry. In this context, an appropriate regulatory framework is essential to optimize the benefits of storage and demand side management within a deregulated environment. Without this, the current regulatory arrangements may present a significant barrier to the introduction of these technologies.
\par
The issue of overlaid networks and their effective interaction is becoming a major issue to understand. This problem is easiest to identify in transport and communication networks by the way in which certain  networks become key hubs.  For example In the UK, the Kings Cross London complex is a major operational and security issue.  Some 15 different international, national and local rail and underground lines are located at the interchange site spread amongst 5 functionally distinct stations.  Similar, but less obvious, problems can arise with overlaying of communication and energy grids.
\par
The resilience of these overlaid structures and the concern of targeted attack makes their functioning critical \cite{Pearson}. It is argued for the Growing ICT dependence in the European grid,  in the context of grids synchronization. Pearson reflects on electricity flows at nearly the speed of light, demanding high-speed decision-making and response times for its management.

\par Secondly, he states that the problem of balancing supply and demand requires instant decision making particularly with the growth of green and inherently unstable sources of supply.
\par Thirdly, he states the flow of alternating current electricity through the grid cannot be straightforwardly controlled by opening or closing a valve in a pipe, or switched like a call on a telephone network.   Electricity is bound by Kirchhoff laws to flow along the network in a free way. The flow levels are determined without regard to capacity as it continually adjusts to its supply demand characteristics. Moreover, a breakdown of an overloaded line results in an instant realignment of the current in the network, and power overload as a result of redistribution of current is  an increasingly likely scenario.  Determining and observing this flow behaviour is a complex task (USA Canada Power System Outage Task Force, 2004).
\par
Some attempts are now in progress to understand the resilience issues of multi-layer networks \cite{Buldyrev}.   The set-up in the model is highly mathematical with a percolation type approach to the onset of breakdown in connections; the results show the suspected weaker resilience of two interlinked networks coming under attack, compared to the resilience in the case where the communication and power network are separate and each operated autonomously.
\subsection{Micro- and nano-scales}
\label{MICRO}
Historically, ICT researchers have paid little attention to availability when searching for materials with specific electronic or magnetic properties. Today,  given the scope and scale of the world's future \emph{Emerging Energy Web} needs, a technology with the capacity to have a significant impact necessarily involves vast quantities of material.
 Unconventional computational schemes, novel physical phenomena, materials and devices (see for example \cite{JRC_Moss,Krohns,Still,Giazotto,Helbing2000,Carbone10,Carbone03}) should be devised to improve the efficiency at the core of the \emph{Emerging Energy Web}. Improvements of the energy performances at the microscopic level would then imply a scaling up at the higher levels of the future \emph{Emerging Energy Web}. Smart appliances,  devices with multiple sensing operativity and remote connectivity are increasingly needed to control the physical layer of the web from the energetic standpoint. Here below we provide just a few examples of the many directions of research for the future control of heating and freezing phenomena  underlying the operation of the Energy Web, trying to pointing out to the multidimensional aspects of the problem, not just to the technological issue.
\par
When we think of scarce natural resources, the availability of fossil fuels, the climate and geopolitical challenges related to the distribution of these resources are certainly  the most immediate (as it has been described earlier in this manuscript). Indeed, we are seeing continuously rising oil prices. However, what has recently become clear is that many other key materials are in danger of becoming unavailable, whereas they are widely needed because of the chemical elements they contain.
The group of materials  are the rare-earth elements, many of whose prices have more than tripled over the past year. As Alexander King, director of the Ames Laboratory in Iowa, US, says in an interview, ``prices will continue to rise. China, who mines 97\% of the global rare-earth metal supply, now requires a significant amount of these elements for its own domestic production. At the same time global demand is also set to rise, making price hikes inevitable". Materials science is only one aspect of the response to these daunting challenges.
\par
With the advent of nano-materials, the lab-on-a-chip engineering is expected to play an increasing role in sustainable technologies for energy conversion, storage and savings. Principal areas of interest are solar cells, batteries and supercapacitors, fuel cells, thermoelectrics, superconductors, more efficient lighting and hydrogen technologies. To achieve an optimal energy conversion and transformation of heat and work at all scales, improvement should begin particularly at the nanoscales. New materials, devices and smart arrays allowing efficient energy transport, generation and detection need to be devised.
In recent years, remarkable progress in the development of materials has led to a continuous increase in the current capacity of high critical temperature superconductor wires and tapes. All these new  generation ICT devices rely on the use of \emph{unfamiliar substances}.
 One can say that the ``\emph{Steel Age}" ended and we are now in the ``\emph{Rare Elements Age"}, when these \emph{unfamiliar substances} will play an increasingly important role in our lives. Extreme variations in the price of some of these elements over the past decade hint at a complex inter-play between rapidly growing demand and limited supply that may become more widespread in the near future. Although the impact of constraints on rare element supplies may reach into every area of the economy, our concern focuses on their effect on new energy technologies.
\par
If new technologies based on a rare energy-critical element were to be widely deployed, widely enough to make a significant contribution to the future \emph{Emerging Energy Web}, quantities of the rare element might be required that exceed present production, perhaps by orders of magnitude. Some of these energy-critical element may not be available in the quantity and at the price necessary to allow large-scale deployment of what might otherwise be a game-changing technology. Scarcity and possible future shortage  of key elements used in modern technology for renewable energy production and transfer is thus a major concern of the coming decades.
The constraints on availability may take many forms. Some potential energy-critical elements as tellurium and rhenium simply are scarce in the earth crust. Rhenium, for example, is 5 times rarer than gold. Others are not so scarce but rarely found in sufficient concentration to be easily mined  \cite{Krohns}.
\par
In summary, the future Energy Web could not survive for long without efficient saving energy schemes even at the smallest scales of its hierarchical structure, where new computational architecture and device design should be devised and employed to slow down Watt counting.

\section{Conclusions}
\label{Conclusions}

The Emerging Energy Web system represents a mayor layer of the FuturICT global platform characterized by many of the challenges that FuturICT wants to address at a global level.  It is a network based technology that will transform our society, the way we interact, live and take decisions regarding the future energy scenario and will require political and institutional transformations. Understanding this Emerging Energy Web will represent the core of the activities of the FuturICT Energy Exploratory.
\par
Thanks to the pervasiveness of ICT design, large quantities of data of different types related to energy sources and use (home, networks, transactions, social interactions) will feed into the `\emph{Planetary Nervous System}'. The modelling of the physical infrastructure will have to be coupled with the economic and social modelling. In particular, agent-based modelling will be used for the representation of the various actors in the decision making processes, in their social interactions, in the interactions with the environment, and in their participation in the market, all this in concerted operation with the other activities of the `\emph{Living Earth Simulator}'. In this context, elements and contribution from social sciences will be essential for representing and understanding the behaviour of actors: prosumers, households, communities (e.g., virtual power plants), institutions and markets. Establishing smart grids requires institutions that support mutual trust and trust in the governance of the framework. The governance framework of smart grids will have to evolve from the current centralised and hierarchical decision making framework towards a polycentric one with many different centres of decision-making at different scales. This is especially valid for the management of failure situations and the related crises. In this context the concept of a `\emph{Global Participatory Platform}' can come into help.
Relevant research questions to be addressed within this research infrastructure, among the others, are:
\begin{itemize}
\item How the multi-dimensional, multi-actors energy system can satisfy the economic, environmental, security and social requirements? ;
\item How the future energy system can  be affected by different threats and risks?
\item  How the `\emph{Planetary Nervous System}' can help in detecting these threats and crises?
\item  How the `\emph{Living Earth Simulator}' can help in modeling the characteristic patterns of these threats and crises?
\item  How the `\emph{Global Participatory Platform}' can help in disseminating this knowledge?
\end {itemize}

 The `\emph{FuturICT Energy Exploratory}' will represent an open accessible framework for citizens, businesses, and organizations enabling to share and explore data and simulations in the energy system scenarios, exploit all potential benefits, while countering disturbances and disruptions, and debate and govern the technical, social and economic implications. In this way, citizen, communities, small businesses will create ecosystems of new energy applications in the wider framework of the `\emph{Innovation Accelerator}' ecosystem, new forms of social and economic participation to explore new future scenarios.
The lessons learned from complex systems, social sciences and advanced ICT tools will jointly come into help in answering all these questions within the FuturICT endeavour.

\section{Acknowledgements}
We are grateful to the anonymous reviewers for many insightful comments.
The publication of this work was partially supported by the European
Union's Seventh Framework Programme (FP7/2007-2013) under grant agreement no.
284709, a Coordination and Support Action in the Information
and Communication Technologies activity area (`FuturICT' FET Flagship Pilot Project).

\end{document}